\documentclass[12pt]{article}
\usepackage{graphicx}

\newcommand{\beq}{\begin{equation}}
\newcommand{\eeq}{\end{equation}}
\newcommand{\beqa}{\begin{eqnarray}}
\newcommand{\eeqa}{\end{eqnarray}}
\newcommand{\non}{\nonumber}
\newcommand{\lab}{\label}

\newcommand{\ket}{\rangle}

\begin{document}

\title{Quantum computation with Kerr-nonlinear photonic crystals}

\author{Hiroo Azuma\thanks{Present address: 2-1-12 MinamiFukunishi-cho,
Oe, NishiKyo-ku, Kyoto-shi, Kyoto 610-1113, Japan,
E-mail: hiroo.azuma@m3.dion.ne.jp}\\
{\small Research Center for Quantum Information Science,}\\
{\small Tamagawa University Research Institute,}\\
{\small 6-1-1 Tamagawa-Gakuen, Machida-shi, Tokyo 194-8610, Japan}
}

\date{\today}

\maketitle

\begin{abstract}
In this paper, we consider a method for implementing a quantum logic gate
with photons whose wave function propagates
in a one-dimensional Kerr-nonlinear photonic crystal.
The photonic crystal causes the incident photons to undergo Bragg reflection
by its periodic structure of dielectric materials
and forms the photonic band structure, namely, the light dispersion relation.
This dispersion relation reduces the group velocity of the wave function
of the photons,
so that it enhances nonlinear interaction of the photons.
(Because variation of the group velocity against the wave vector is very steep,
we have to tune up the wavelength of injected photons precisely, however.)
If the photonic crystal includes layers of a Kerr medium,
we can rotate the phase of the wave function of the incident photons
by a large angle efficiently.
We show that we can construct the nonlinear sign-shift (NS) gate proposed
by Knill, Laflamme, and Milburn (KLM) by this method.
Thus, we can construct the conditional sign-flip gate for two qubits,
which is crucial for quantum computation.
Our NS gate works with probability unity in principle
while KLM's original one is a nondeterministic gate conditioned
on the detection of an auxiliary photon.
\end{abstract}

\section{Introduction}
Since Shor's quantum algorithms for prime factorization and discrete logarithms appeared,
many researchers have been devoting their efforts
to realization of quantum computation
in their laboratory \cite{Shor}.
To implement a quantum computer, we have to prepare qubits,
which are two-state systems,
and quantum logic gates, which apply unitary transformations to qubits.
Generally speaking, we can apply an arbitrary transformation of $U(2)$ group to a qubit at ease
no matter which physical system we choose as the qubit.
However, almost all the researchers think implementation of a two-qubit gate
to be very difficult
because the two-qubit gate has to generate entanglement between two local systems.
Moreover, it is shown that we can construct any unitary transformation
applied to qubits
from $U(2)$ transformations and a certain two-qubit gate,
such as the controlled-NOT gate or the conditional sign-flip gate
\cite{Nielsen-Chuang,Sleator-Weinfurter,
Barenco-Bennett-Cleve-DiVincenzo-Margolus-Shor-Sleator-Smolin-Weinfurter}.
From the above reasons, theoretical and experimental physicists aim
for realizing the two-qubit gate.

Among proposals made for implementing quantum computation,
for example realization of qubits by cold trapped ions interacting with laser beams
\cite{Cirac-Zoller,Monroe-Meekhof-King-Itano-Wineland},
polarized photons in the cavity quantum electrodynamics system
\cite{Turchette-Hood-Lange-Mabuchi-Kimble},
nuclear spins of molecules under nuclear magnetic resonance \cite{Gershenfeld-Chuang},
and so on,
the scheme of Knill, Laflamme, and Milburn (KLM) is very unique
\cite{Knill-Laflamme-Milburn,Ralph-White-Munro-Milburn}.
They show a method for constructing the conditional sign-flip gate
from linear optical elements
(single photon sources, beamsplitters, and photodetectors).
Because many researchers believe that only nonlinear interaction
between two qubits can generate
quantum correlation,
KLM's scheme that does not require any nonlinear devices seems
to be novel and attractive.
In KLM's method, Bose-Einstein statistics of photons play an important role,
so that we can observe bunching of photons in outputs of beamsplitters.
Recent progress about KLM's linear optical quantum computing is reviewed in
Ref.~\cite{Kok-Munro-Nemoto-Ralph-Dowling-Milburn}.

To implement the conditional sign-flip gate,
KLM use the nonlinear sign-shift (NS) gate,
which applies the following transformation to a superposition
of the number states of photons
$|n\ket$ for $n=0,1,2$:
$\alpha|0\ket+\beta|1\ket+\gamma|2\ket
\rightarrow
\alpha|0\ket+\beta|1\ket-\gamma|2\ket$
(with probability $1/4$).
The NS gate is essential for KLM's scheme,
and they construct this gate only  from passive linear optics.
Although KLM's NS gate is a nondeterministic gate that works with probability $1/4$,
we can detect its false event by measuring an auxiliary photon.

In this paper, we discuss the method for implementing the NS gate
with a one-dimensional Kerr-nonlinear photonic crystal.

A photonic crystal is a periodic structure of dielectric materials
whose dielectric constants are different from each other.
If a wavelength of a propagating electromagnetic wave is comparable
with the period of the crystal,
Bragg reflection occurs and the photonic band gap (the light dispersion relation)
is formed \cite{Yablonovitch-PRL,John,Yablonovitch-JOSAB}.
This dispersion relation reduces the group velocity of the electromagnetic field
of incident photons,
and thus it enhances nonlinear interaction of the photons.
If we construct the photonic crystal from alternating layers of a Kerr medium
and a medium with linear polarization
(a medium with $\chi^{(2)}=0$ and $\chi^{(3)}\neq 0$
and a medium with $\chi^{(2)}=\chi^{(3)}=0$,
where $\chi^{(n)}$ is the $n$th-order nonlinear electric susceptibility)
as a one-dimensional periodic structure,
we can rotate the phase of the wave function of the incident photons
by a large angle efficiently.
(In general, for almost all the dielectric media,
Kerr nonlinearity is too weak to rotate the phase of wave function of photons
by a certain angle.)
Using this method, we realize KLM's NS gate.
In our implementation,
the NS gate works with probability unity in principle.

Someone may make an objection to our proposal
because we are going to introduce a nonlinear device
into KLM's scheme that aims to construct a quantum computer
only with passive linear optics.
However, the author thinks that our proposal is a practical compromise
because the Kerr medium is a well-studied material in the field of optics
and rapid progress has been made in experimental studies of photonic crystals.
The idea of making use of the Kerr medium for the conditional sign-flip gate
can be found in Ref.~\cite{Chuang-Yamamoto}.
Inoue and Aoyagi show experimental evidence of enhancement of the nonlinearity in
a Kerr-nonlinear photonic crystal in Ref.~\cite{Inoue-Aoyagi}.
Numerical investigations of a one-dimensional Kerr-nonlinear photonic crystal
are made in Refs.~\cite{Scalora-Dowling-Bowden-Bloemer,Huttunen-Torma}.
Some research groups have recently proposed optical quantum computation schemes
using weak cross-Kerr nonlinearities and strong probe coherent fields
\cite{Fiurasek,Nemoto,Munro,Lee,Jeong}.
Their proposals and our method use optical nonlinearities in common with each other.

This paper is organized as follows:
In the rest of this section, we give a brief review of KLM's scheme.
In Sec.~\ref{section-outline-NS-gate},
we outline the NS gate realized by the Kerr-nonlinear photonic crystal.
And then, we estimate the time-of-flight that the injected photons need
for obtaining a certain angle of the phase rotation caused by nonlinear interaction
in the homogeneous Kerr medium.
In Sec.~\ref{section-Dispersion-relation-Kerr-nonlinear-photonic-crystal},
we investigate the photonic band structure in the photonic crystal
and evaluate the group velocity of the photons.
From these results, we give a concrete example of a design for the photonic crystal
that realizes the NS gate with giving numerical values of the period of the crystal,
the thickness and the dielectric constant of each layer, and so on.
In Sec.~\ref{section-discussion}, we give discussions.
In Appendix~\ref{section-third-order-nonlinear-susceptibility-GaAs/GaAlAs},
we explain the third-order nonlinear susceptibility $\chi^{(3)}$
and the nonlinear refraction coefficient $n_{2}$.
In Appendix~\ref{section-wave-equation-effective-Hamiltonian-nonlinear-dielectric-medium},
we derive the wave equation and the effective Hamiltonian of the electromagnetic field
in the nonlinear dielectric medium.

Here, we sketch out KLM's quantum circuit for implementing the conditional sign-flip gate.
First, we construct a qubit from a pair of optical paths.
The optical path (mode) forms a physical system which takes
a superposition of the number state $|n\ket$ for $n=0,1,2,...$,
where $n$ is the number of photons on the path.
$|0\rangle_{x1}\otimes|1\rangle_{x2}$ is a state where modes $x1$ and $x2$ have zero and one photons,
respectively, and we regard it as a logical ket vector $|\bar{0}\rangle_{x}$.
We regard $|1\ket_{x1}\otimes|0\ket_{x2}$ as a logical ket vector $|\bar{1}\ket_{x}$, similarly.
And we describe an arbitrary state of a qubit as
$|\phi\ket_{x}=\alpha|\bar{0}\ket_{x}+\beta|\bar{1}\ket_{x}$
for
$|\alpha|^{2}+|\beta|^{2}=1$.
(This construction of a qubit is called the dual-rail qubit representation
\cite{Chuang-Yamamoto}.)

Second, we define the NS operation as follows:
\beqa
&&
|\psi\ket=\alpha|0\ket+\beta|1\ket+\gamma|2\ket \non \\
&&\quad\quad
\rightarrow
|\psi'\ket=\alpha|0\ket+\beta|1\ket-\gamma|2\ket.
\lab{definition-NS-operation}
\eeqa
We pay attention to the fact that the NS operation does not work on a qubit
but on a superposition of the number states of a single mode,
$|0\ket$, $|1\ket$, and $|2\ket$.

\begin{figure}
\begin{center}
\includegraphics[scale=1.0]{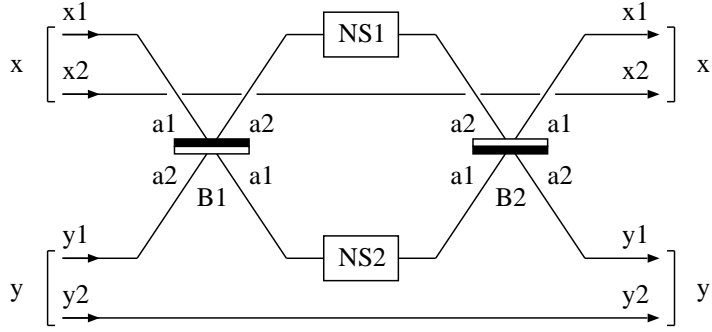}
\end{center}
\caption{Implementation of the conditional sign-flip gate
with the NS operations.
Qubits $x$ and $y$ consist of a pair of modes $x1$ and $x2$ and a pair of modes $y1$ and $y2$,
respectively.
Symbols $B1$ and $B2$ represent beamsplitters.
Symbols $NS1$ and $NS2$ represent the NS gates.
Photons travel from left to right in this network.}
\lab{CSFgate}
\end{figure}

Third, we construct the conditional sign-flip gate from the NS operation as shown
in Fig.~\ref{CSFgate}.
An optical network drawn in Fig.~\ref{CSFgate} works as the conditional sign-flip gate,
whose operation is given by
\beqa
&&|\bar{j}\ket_{x}\otimes|\bar{k}\ket_{y}
\rightarrow
(-1)^{jk}
|\bar{j}\ket_{x}\otimes|\bar{k}\ket_{y} \non \\
&&\quad\quad
\mbox{for $j,k\in\{0,1\}$}.
\eeqa
Let us confirm the function of this network.
In Fig.~\ref{CSFgate},
symbols $B1$ and $B2$ represent beamsplitters,
which transform the incident number states of modes $a1$ and $a2$ as follows:
\beqa
&&
|n\ket_{a1}|m\ket_{a2}
=
\frac{1}{\sqrt{n!m!}}(a_{1}^{\dagger})^{n}(a_{2}^{\dagger})^{m}|0\ket_{a1}|0\ket_{a2} \non \\
&\rightarrow&
\frac{1}{\sqrt{n!m!}}
[\frac{1}{\sqrt{2}}(a_{1}^{\dagger}+a_{2}^{\dagger})]^{n}
[\frac{1}{\sqrt{2}}(a_{1}^{\dagger}-a_{2}^{\dagger})]^{m}
|0\ket_{a1}|0\ket_{a2} \non \\
&&\quad\quad
\mbox{for $n,m\in\{0,1,2,...\}$},
\eeqa
where $a_{1}^{\dagger}$ and $a_{2}^{\dagger}$ are creation operators of modes $a1$ and $a2$
respectively,
and their commutation relations are given by
$[a_{j},a_{k}^{\dagger}]=\delta_{jk}$
and
$[a_{j},a_{k}]=[a_{j}^{\dagger},a_{k}^{\dagger}]=0$
for
$j,k\in\{1,2\}$.
The beamsplitters $B1$ and $B2$ replace $a_{1}^{\dagger}$ and $a_{2}^{\dagger}$
with
$(1/\sqrt{2})(a_{1}^{\dagger}+a_{2}^{\dagger})$
and
$(1/\sqrt{2})(a_{1}^{\dagger}-a_{2}^{\dagger})$, respectively.
We give attention to the fact that
$B2$ applies an inverse transformation of $B1$.
Symbols $NS1$ and $NS2$ represent the NS gates that apply the operation given
in Eq.~(\ref{definition-NS-operation})
to the modes $a1$ and $a2$, respectively.

If we put a superposition of
$|\bar{0}\ket_{x}|\bar{0}\ket_{y}$,
$|\bar{0}\ket_{x}|\bar{1}\ket_{y}$,
and
$|\bar{1}\ket_{x}|\bar{0}\ket_{y}$
into the left side of the network shown in Fig.~\ref{CSFgate},
the network leaves it untouched and returns it as an output
from the right side of the network.
However, if we put a state
$|\bar{1}\ket_{x}|\bar{1}\ket_{y}
=
|1\ket_{x1}|0\ket_{x2}|1\ket_{y1}|0\ket_{y2}$
into the network,
the following transformation is applied to the modes $x1$ and $y1$:
\beqa
|1\ket_{x1}|1\ket_{y1}
&\stackrel{B1}{\longrightarrow}&
\frac{1}{\sqrt{2}}
(|2\ket_{a1}|0\ket_{a2}-|0\ket_{a1}|2\ket_{a2}) \non \\
&\stackrel{NS1,NS2}{\longrightarrow}&
-\frac{1}{\sqrt{2}}
(|2\ket_{a1}|0\ket_{a2}-|0\ket_{a1}|2\ket_{a2}) \non \\
&\stackrel{B2}{\longrightarrow}&
-|1\ket_{x1}|1\ket_{y1}.
\eeqa
Thus, we obtain
$-|\bar{1}\ket_{x}|\bar{1}\ket_{y}
=
-|1\ket_{x1}|0\ket_{x2}|1\ket_{y1}|0\ket_{y2}$
as an output for the input state $|\bar{1}\ket_{x}|\bar{1}\ket_{y}$.

Therefore, we confirm that the network shown in Fig.~\ref{CSFgate}
realizes the conditional sign-flip gate.
Moreover, we notice that implementing the NS gate is essential
for the network in Fig.~\ref{CSFgate}.

\section{An outline of the NS gate}
\lab{section-outline-NS-gate}
In this section, we explain the NS gate,
which is constructed from a one-dimensional Kerr-nonlinear photonic crystal.
As shown in Fig.~\ref{NSgate},
we inject photons whose wave function is given by $|\psi\ket$
(a superposition of $|0\ket$, $|1\ket$, and $|2\ket$)
into the photonic crystal.
The photonic crystal is composed of alternating layers of materials $A$ and $B$,
and they form a one-dimensional periodic structure with a period $(l_{A}+l_{B})$.
The material $A$ is a linear dielectric medium
whose dielectric constant is given by $\epsilon_{A}$,
and the material $B$ is a Kerr medium
whose dielectric constant and third-order nonlinear susceptibility
for optical Kerr effect are given by $\epsilon_{B}$ and $\chi_{B}^{(3)}$, respectively.
While the wave function of the photons travels through the photonic crystal,
it undergoes Bragg reflection and its group velocity decreases.
At the same time, the phase of the wave function is rotated by the optical Kerr effect
induced by the layers of he Kerr medium $B$.
Because the slow group velocity enhances the Kerr nonlinearity,
we can expect the phase to rotate by a large angle efficiently.

\begin{figure}
\begin{center}
\includegraphics[scale=1.0]{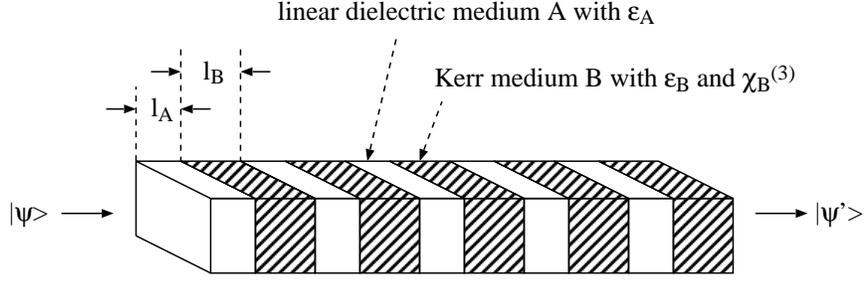}
\end{center}
\caption{An outline of the NS gate constructed
from a one-dimensional Kerr-nonlinear photonic crystal.
The photonic crystal is composed of alternating layers of materials $A$ and $B$
with thicknesses $l_{A}$ and $l_{B}$, respectively.
[Thus, a period of the crystal is given by $(l_{A}+l_{B})$.]
The material $A$ is a linear dielectric medium
whose dielectric constant is given by $\epsilon_{A}$,
and the material $B$ is a Kerr medium whose dielectric constant and
third-order nonlinear susceptibility for optical Kerr effect are given by $\epsilon_{B}$
and $\chi_{B}^{(3)}$, respectively.
A wave function of injected photons $|\psi\ket$ propagates
through the photonic crystal with nonlinear phase rotation of the optical Kerr effect
enhanced by the slow group velocity,
and it goes out of the photonic crystal as $|\psi'\ket$.}
\lab{NSgate}
\end{figure}

Here, we estimate the rotational angle of the phase caused by the optical Kerr effect
in a homogeneous Kerr medium by way of trial.
(We do not investigate the phase rotation
in a one-dimensional Kerr-nonlinear photonic crystal here.
We evaluate the phase rotation caused by the photonic crystal
in Sec.~\ref{section-Dispersion-relation-Kerr-nonlinear-photonic-crystal}.)
According to the Refs.~\cite{Drummond-Walls,Mandel-Wolf},
we have the effective Hamiltonian of photons in the homogeneous Kerr medium
with $\epsilon$ and $\chi^{(3)}$ as
\beq
\mathcal{H}
\simeq
\hbar\omega
(a^{\dagger}a+\frac{1}{2})
+\hbar\chi(a^{\dagger})^{2}a^{2},
\lab{effective-Hamiltonian-Kerr-medium}
\eeq
and
\beq
\chi
=
\frac{9}{8}
\frac{\hbar\omega^{2}\chi^{(3)}}{\epsilon^{2}Sd},
\lab{definition-Kerr-constant}
\eeq
where
$\omega$ is a circular frequency of the photons,
$S$ represents the cross-sectional area of a device made of the Kerr medium,
and $d$ represents the width of the wave-packet of the photons in the medium.
Thus, $Sd$ is the volume for the quantization of photons.
[While $\chi^{(3)}$ is given by a fourth-rank tensor as
$\mbox{\boldmath $P$}
=
\chi^{(1)}:\mbox{\boldmath $E$}
+
\chi^{(3)}:\mbox{\boldmath $E$}\mbox{\boldmath $E$}\mbox{\boldmath $E$}$ in general,
$\chi^{(3)}$ in Eq.~(\ref{definition-Kerr-constant}) represents
a certain component of the fourth-rank tensor.
In Appendix~\ref{section-third-order-nonlinear-susceptibility-GaAs/GaAlAs},
we give a brief explanation about $\chi^{(3)}$
and the nonlinear refraction coefficient $n_{2}$,
which is observed in an experiment direct.
In Appendix~\ref{section-wave-equation-effective-Hamiltonian-nonlinear-dielectric-medium},
we show derivation of the effective Hamiltonian given by
Eqs.~(\ref{effective-Hamiltonian-Kerr-medium})
and (\ref{definition-Kerr-constant}).]

To derive the effective Hamiltonian given by
Eqs.~(\ref{effective-Hamiltonian-Kerr-medium})
and (\ref{definition-Kerr-constant}),
we carry out the following process.
First of all, we perform the quantization of the free electromagnetic field
whose Hamiltonian does not include $\chi^{(3)}$.
Next, regarding $\chi^{(3)}$ as an expansion parameter,
we evaluate the first-order perturbed energy,
which we can consider to be the effective Hamiltonian.
(This process is explained in
Appendix~\ref{section-wave-equation-effective-Hamiltonian-nonlinear-dielectric-medium}.)
In the derivation of the Hamiltonian given in
Eqs.~(\ref{effective-Hamiltonian-Kerr-medium})
and (\ref{definition-Kerr-constant}),
we assume that photons travel in the homogeneous Kerr medium.
However, the system that we discuss in this paper is photons
propagating through the photonic crystal.
Thus, we have to consider the wave function of photons
which extends over many layers of the photonic crystal.
Therefore, the Hamiltonian given
by Eqs.~(\ref{effective-Hamiltonian-Kerr-medium})
and (\ref{definition-Kerr-constant}) offers us
a rough picture of the phase rotation.
(Moreover, although we discuss the phase rotation by quantum mechanics in this section,
we derive the photonic band gap by classical theory of electrodynamics
in Sec.~\ref{section-Dispersion-relation-Kerr-nonlinear-photonic-crystal}.
Hence, our analysis of the quantum gate is semi-classical.)

According to Eq.~(\ref{effective-Hamiltonian-Kerr-medium}),
the number state of the photons $|n\ket$ for $n=0,1,2,...$ obtains the phase rotation
$\exp[i\chi n(n-1)t]$,
where $t$ is photons' time-of-flight in the crystal.
[We remember $(a^{\dagger})^{2}a^{2}|n\ket=n(n-1)|n\ket$.
Because the first term of Eq.~(\ref{effective-Hamiltonian-Kerr-medium})
(namely, the unperturbed Hamiltonian) gives only the background phase rotation
$\exp\{i\omega[n+(1/2)]t\}$ for $|n\ket$,
we can neglect it.]
Thus, the Kerr medium transforms the wave function of the traveling photons as
$\alpha|0\ket+\beta|1\ket+\gamma|2\ket
\rightarrow
\alpha|0\ket+\beta|1\ket+e^{i2\chi t}\gamma|2\ket$.
If we let $|\chi t|$ be equal to $\pi/2$,
we can use this photonic crystal as the NS gate.

In Refs.~\cite{Miller-Chemla-Eilenberger-Smith-Gossard-Tsang,
Miller-Chemla-Eilenberger-Smith-Gossard-Wiegmann,
Chemla-Miller-Smith},
nonlinear optical properties of GaAs/GaAlAs multiple quantum well (MQW) material are studied.
In Ref.~\cite{Miller-Chemla-Eilenberger-Smith-Gossard-Wiegmann},
Miller {\it et al}. prepare a sample that consists of
84 periods of 144-{\AA} GaAs and 102-{\AA} GaAs/$\mbox{Ga}_{0.7}\mbox{Al}_{0.3}$As
and obtain the nonlinear refraction coefficient
$n_{2}\sim 1.2\times 10^{-4}$ [$\mbox{cm}^{2}$/W]
at wavelength ($\sim 850$ nm).
The dielectric constants of GaAs and GaAlAs are given by $\epsilon=13.0\times \epsilon_{0}$,
where $\epsilon_{0}$($=8.85\times 10^{-12}$ [F/m]) is the dielectric constant of vacuum.
From the result of $n_{2}\sim 1.2\times 10^{-4}$ [$\mbox{cm}^{2}$/W],
the third-order nonlinear susceptibility of GaAs/GaAlAs MQW structure
is given by
$\chi^{(3)}=2.97\times 10^{-2}$ (esu)$=3.67\times 10^{-21}$ [m$\cdot$C/$\mbox{V}^{3}$] (SI unit).
This $|\chi^{(3)}|$ is much larger than those of other materials.
[The relation between $n_{2}$ and $\chi^{(3)}$
is explained in Appendix~\ref{section-third-order-nonlinear-susceptibility-GaAs/GaAlAs}
\cite{Yariv}.]

Let us estimate $t=\pi/(2\chi)$,
the time-of-flight for realizing the conditional sign-flip operation, numerically.
We assume that the wavelength of the incident photons is given by
$\lambda=8.47\times 10^{-7}$ [m]
[$\omega=2\pi(c/\lambda)=2.23\times 10^{15}$ [rad/s],
$\nu=c/\lambda=3.54\times 10^{14}$ [1/s]]
which is slightly out of tune from 850 nm.
(We use $c=3.00\times 10^{8}$ [m/s].)
We also assume that the incident photons form a wave-packet
whose cross-sectional area and width
in vacuum are given by
$S=1.0\times 10^{-8}$ [$\mbox{m}^{2}$]
(a square with the side-length $0.1$ mm)
and
$d_{0}=4.24\times 10^{-6}$ [m]
(five times as long as the wavelength), respectively.
This wave-packet corresponds with a $14.1$ femto-second pulse.
Because the width of the wave-packet is shortened in the medium as
$d=\sqrt{(\epsilon_{0}\mu_{0})/(\epsilon\mu)}d_{0}$,
where $\mu_{0}$ and $\mu$ represent the magnetic permeabilities of vacuum
and the material respectively,
we obtain $d=1.18\times 10^{-6}$ [m] in GaAs/GaAlAs.
(We use the fact $\mu\simeq\mu_{0}$ for GaAs and GaAlAs.)
The Planck constant is given by
$\hbar=h/(2\pi)=1.05\times 10^{-34}$ [J$\cdot$s].
Substituting the above numerical values into Eq.~(\ref{definition-Kerr-constant}),
we obtain $\chi=1.38\times 10^{10}$ [1/s].
Thus, we need $t=\pi/(2\chi)=1.14\times 10^{-10}$ [s] for the time-of-flight
in GaAs/GaAlAs MQW material.
To obtain the time-of-flight $t=1.14\times 10^{-10}$ [s] in the GaAs/GaAlAs MQW material,
we have to prepare a device whose length is equal to
$L=\sqrt{(\epsilon_{0}\mu_{0})/(\epsilon\mu)}ct=9.49\times 10^{-3}$ [m].
(In this evaluation, we assume that the velocity of the photons in the Kerr medium is not affected by
the term of $\chi$.
This is because the contribution of the term of $\chi$ is much smaller than the Hamiltonian
of the free photons.
We notice that the term of $\chi$ in the Hamiltonian increases in proportion to the square of the number
of the photons and we consider the case where the number of the photons is equal to two at most now.
Details of this discussion is given
in Sec.~\ref{section-Dispersion-relation-Kerr-nonlinear-photonic-crystal}.)

In the above estimation, the volume for the quantization of photons $Sd$ is crucial.
We assume $S=1.0\times 10^{-8}$ [$\mbox{m}^{2}$] and this quantity corresponds
with the cross-sectional area of the device of the GaAs/GaAlAs MQW material.
To prevent the wave function of photons from tunneling outside of the box for quantization,
we have to wrap up the device of GaAs/GaAlAs in a material whose refractive index is larger
than refractive indexes of both GaAs and GaAlAs.
This means that we have to make a cavity of the GaAs/GaAlAs MQW material
for confining photons inside,
and the cross-sectional area and the length of the cavity is given by
$S=1.0\times 10^{-8}$ [$\mbox{m}^{2}$] and $L=9.49\times 10^{-3}$ [m],
respectively.

The author cannot judge whether this requirement is feasible for realizing in the laboratory.
Because the ratio between two side-lengths of the cavity is very large,
it seems to be difficult to fabricate this device.
To overcome this problem,
we reduce the group velocity of photons in the Kerr medium by the dispersion relation
induced by the photonic crystal.

\section{Dispersion relation of the Kerr-nonlinear photonic crystal}
\lab{section-Dispersion-relation-Kerr-nonlinear-photonic-crystal}
In this section, we investigate the photonic band gap (the dispersion relation) induced
in the photonic crystal of Fig.~\ref{NSgate}.
And then, we derive the group velocity of incident photons from this dispersion relation.

The wave equation of the electromagnetic field in a one-dimensional Kerr-nonlinear photonic crystal
is given by
\beqa
&&
\frac{\partial^{2}}{\partial z^{2}}E(z,t)
-
\mu_{0}\epsilon(z)\frac{\partial^{2}}{\partial t^{2}}E(z,t) \non \\
&&\quad\quad
-
\mu_{0}\chi^{(3)}(z)\frac{\partial^{2}}{\partial t^{2}}E(z,t)^{3}=0,
\lab{wave-equation-electric-field-in-nonlinear-medium}
\eeqa
\beq
\frac{\partial}{\partial z}E(z,t)
=
-\frac{\partial}{\partial t}B(z,t),
\lab{wave-equation-magnetic-field-in-nonlinear-medium}
\eeq
where
\beq
\epsilon(z+l_{A}+l_{B})
=\epsilon(z)
\quad\quad
\mbox{for $-\infty <x<\infty$},
\lab{periodicity-dielectric-constant}
\eeq
\beq
\epsilon(z)=
\left\{
\begin{array}{lll}
\epsilon_{A}(>0) & \quad & 0\leq z<l_{A} \\
\epsilon_{B}(>0) & \quad & l_{A}\leq z<l_{A}+l_{B}
\end{array}
\right.,
\lab{definition-dielectric-constant-Kronig-Penney}
\eeq
\beq
\chi^{(3)}(z+l_{A}+l_{B})
=\chi^{(3)}(z)
\quad\quad
\mbox{for $-\infty <x<\infty$},
\eeq
\beq
\chi^{(3)}(z)=
\left\{
\begin{array}{lll}
0          & \quad & 0\leq z<l_{A} \\
\chi^{(3)} & \quad & l_{A}\leq z<l_{A}+l_{B}
\end{array}
\right..
\eeq
In Eq.~(\ref{wave-equation-electric-field-in-nonlinear-medium}),
we assume that the photonic crystal is periodic in the $z$-direction
and uniform in the $x$-$y$ plane,
and the electromagnetic wave propagates in the positive $z$-direction.
Derivation of Eqs.~(\ref{wave-equation-electric-field-in-nonlinear-medium})
and (\ref{wave-equation-magnetic-field-in-nonlinear-medium}) is shown
in Appendix~\ref{section-wave-equation-effective-Hamiltonian-nonlinear-dielectric-medium}.

Here, we assume that the medium $B$ is the GaAs/GaAlAs MQW material
and all physical quantities are given in Sec.~\ref{section-outline-NS-gate}.
We can estimate the contribution made by the term
$\mu_{0}\chi^{(3)}(z)(\partial^{2}/\partial t^{2})E(z,t)^{3}$
of Eq.~(\ref{wave-equation-electric-field-in-nonlinear-medium}) as follows:
Because of Eq.~(\ref{definition-Kerr-constant}) and
Eqs.~(\ref{product-form-E}), (\ref{special-solutions-u-l}),
(\ref{canonical-variable-p-l})
in Appendix~\ref{section-wave-equation-effective-Hamiltonian-nonlinear-dielectric-medium},
we obtain the relation
\beq
\chi^{(3)}E^{2}
\sim
\chi^{(3)}
\frac{\hbar\omega}{\epsilon Sd}
=
\epsilon
\frac{8}{9}\frac{\chi}{\omega}.
\eeq
(We pay attention to the fact that the number of photons in the wave-packet is
equal to two at most.)
And, using $\chi=1.38\times 10^{10}$ [1/s] and $\omega=2.23\times 10^{15}$ [rad/s],
we obtain
\beq
\frac{8}{9}\frac{\chi}{\omega}
=
5.50\times 10^{-6}.
\eeq
Thus, we arrive at
\beq
\chi^{(3)}E^{2}\ll\epsilon.
\eeq

Hence, we can neglect the term that include $\chi^{(3)}$
in Eq.~(\ref{wave-equation-electric-field-in-nonlinear-medium})
for evaluating the photonic band gap.
Moreover, we take monochromatic light as an approximation of the injected pulse into the photonic
crystal,
so that we only need to consider a stationary solution,
\beq
E(z,t)=E(z)e^{-i\omega t}.
\eeq
(In Sec.~\ref{section-outline-NS-gate}, we assume that the signal injected to the photonic crystal
is a femto-second pulse of the typical length $5\lambda$.
Propagation of ultrashort pulses in the Kerr-nonlinear photonic crystal is considered
in Ref.~\cite{Scalora-Dowling-Bowden-Bloemer}.
Scalora {\it et al}. indicate that the nonlinearity causes various effects to the pulse
near the band edge.
However, because we need hard calculations for analyzing the behavior of the ultrashort pulse
in the photonic crystal
and it is beyond the purpose of this paper,
we neglect these effects for simplicity.
Thus, we adopt the approximation by the monochromatic light.)
Thus, the wave equation that we have to solve is written down as
\beq
[\frac{\partial^{2}}{\partial z^{2}}
+\frac{\epsilon(z)}{\epsilon_{0}}(\frac{\omega}{c})^{2}]
E(z)=0,
\lab{wave-equation-Kronig-Penney}
\eeq
where Eqs.~(\ref{periodicity-dielectric-constant})
and (\ref{definition-dielectric-constant-Kronig-Penney}) are assumed.
This problem is known as Kronig-Penney model and we can derive its exact solutions
\cite{Kronig-Penney,Kittel}.

Let us solve Eqs.~(\ref{periodicity-dielectric-constant}),
(\ref{definition-dielectric-constant-Kronig-Penney}),
and (\ref{wave-equation-Kronig-Penney}).
Solutions of the region I ($0<z<l_{A}$) and the region II ($l_{A}<z<l_{A}+l_{B}$)
for Eqs.~(\ref{definition-dielectric-constant-Kronig-Penney})
and (\ref{wave-equation-Kronig-Penney})
are given by
\beqa
E_{j}(z)
&=&C_{j+}\exp(iK_{j}z)
+C_{j-}\exp(-iK_{j}z) \non \\
&&
\quad\quad
\mbox{for $j=\mbox{I},\mbox{II}$,}
\lab{solution-electric-field-region-I-II}
\eeqa
where
\beq
K_{\mbox{\scriptsize I}}=\frac{\omega}{c}\sqrt{\frac{\epsilon_{A}}{\epsilon_{0}}},
\quad\quad
K_{\mbox{\scriptsize II}}=\frac{\omega}{c}\sqrt{\frac{\epsilon_{B}}{\epsilon_{0}}}.
\lab{wave-vector-I-II}
\eeq

At points where a finite potential step exists,
the above solution must be continuous together with their derivatives
(that is, the continuity conditions).
Moreover, because $\epsilon(z)$ is periodic,
we can apply Bloch's theorem to the continuity conditions of the solution.

\bigskip

\noindent
{\bf Bloch's Theorem}:
Suppose $E_{k}(z)$ is a solution of
Eqs.~(\ref{periodicity-dielectric-constant}) and (\ref{wave-equation-Kronig-Penney}).
Then, $E_{k}(z)$ satisfies the following relations.
\beq
E_{k}(z)=e^{ikz}u_{k}(z),
\eeq
\beq
u_{k}(z+l_{A}+l_{B})
=u_{k}(z)
\quad\quad
\mbox{for $-\infty <z<\infty$},
\eeq
\beq
-\frac{\pi}{l_{A}+l_{B}}\leq k\leq\frac{\pi}{l_{A}+l_{B}}.
\eeq

\bigskip

Thus, we can require the continuity conditions
to $E_{\mbox{\scriptsize I}}(z)$ and $E_{\mbox{\scriptsize II}}(z)$
as follows:
\beq
E_{\mbox{\scriptsize I}}(0)
=
E_{\mbox{\scriptsize II}}(0),
\eeq
\beq
\left.\frac{d}{dz}E_{\mbox{\scriptsize I}}\right|_{z=0}
=
\left.\frac{d}{dz}E_{\mbox{\scriptsize II}}\right|_{z=0},
\eeq
\beq
E_{\mbox{\scriptsize I}}(l_{A})
=
e^{ik(l_{A}+l_{B})}E_{\mbox{\scriptsize II}}(-l_{B}),
\eeq
\beq
\left.\frac{d}{dz}E_{\mbox{\scriptsize I}}\right|_{z=l_{A}}
=
e^{ik(l_{A}+l_{B})}\left.\frac{d}{dz}E_{\mbox{\scriptsize II}}\right|_{z=-l_{B}}.
\eeq
The above four equations can be rewritten in the form
\beq
M
\left(
\begin{array}{c}
C_{\mbox{\scriptsize I}+}\\
C_{\mbox{\scriptsize I}-}\\
C_{\mbox{\scriptsize II}+}\\
C_{\mbox{\scriptsize II}-}
\end{array}
\right)
=0,
\lab{Matrix-equation-continuity-condition}
\eeq
where
\beq
M=
\left(
\begin{array}{cccc}
1 & 1 & -1 & -1 \\
K_{\mbox{\scriptsize I}} & -K_{\mbox{\scriptsize I}}
& -K_{\mbox{\scriptsize II}} & K_{\mbox{\scriptsize II}}\\
P & 1/P & -Q/R & -QR \\
K_{\mbox{\scriptsize I}}P & -K_{\mbox{\scriptsize I}}/P
& -K_{\mbox{\scriptsize II}}Q/R & K_{\mbox{\scriptsize II}}QR
\end{array}
\right),
\lab{definition-matrix-M}
\eeq
\beq
P=\exp(iK_{\mbox{\scriptsize I}}l_{A}),
\eeq
\beq
Q=\exp[ik(l_{A}+l_{B})],
\eeq
\beq
R=\exp(iK_{\mbox{\scriptsize II}}l_{B}).
\eeq
From Eq.~(\ref{Matrix-equation-continuity-condition}),
we obtain $\det M=0$,
and it implies
\beqa
&&
\cos[(l_{A}+l_{B})k]
-\cos(l_{A}K_{\mbox{\scriptsize I}})\cos(l_{B}K_{\mbox{\scriptsize II}}) \non \\
&&\quad\quad
+
\frac{K_{\mbox{\scriptsize I}}^{2}+K_{\mbox{\scriptsize II}}^{2}}
{2K_{\mbox{\scriptsize I}}K_{\mbox{\scriptsize II}}}
\sin(l_{A}K_{\mbox{\scriptsize I}})\sin(l_{B}K_{\mbox{\scriptsize II}})=0.
\lab{equation-light-dispersion-relation}
\eeqa
From Eq.~(\ref{equation-light-dispersion-relation}),
we can obtain the light dispersion relation $\omega=\omega(k)$.

\begin{figure}
\begin{center}
\includegraphics[scale=1.0]{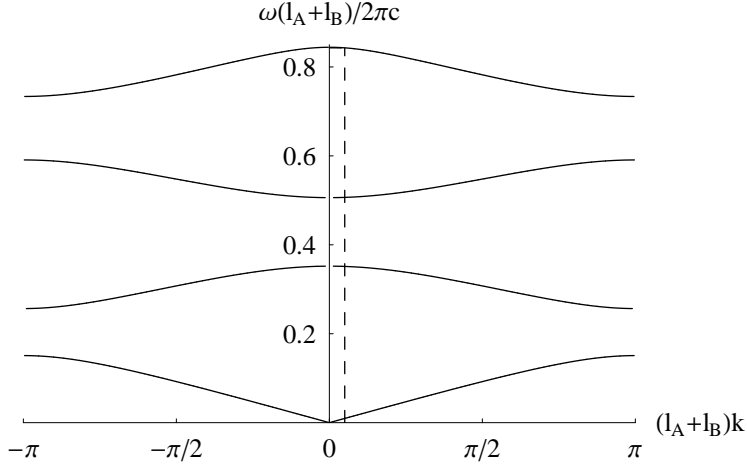}
\end{center}
\caption{The dispersion relation of the photonic crystal
with $l_{A}=l_{B}=3.57\times 10^{-7}$ [m],
$\epsilon_{A}/\epsilon_{0}=1.0$, and $\epsilon_{B}/\epsilon_{0}=13.0$.
Broken lines indicates a point
where $\omega(l_{A}+l_{B})/2\pi c=0.843$
(that is, $\omega=2.23\times 10^{15}$ [rad/s] and $\lambda=8.47\times 10^{-7}$ [m])
and $(l_{A}+l_{B})k=0.158$.
This point belongs to the fourth conduction band.}
\lab{Ph-BandGap}
\end{figure}

\begin{figure}
\begin{center}
\includegraphics[scale=1.0]{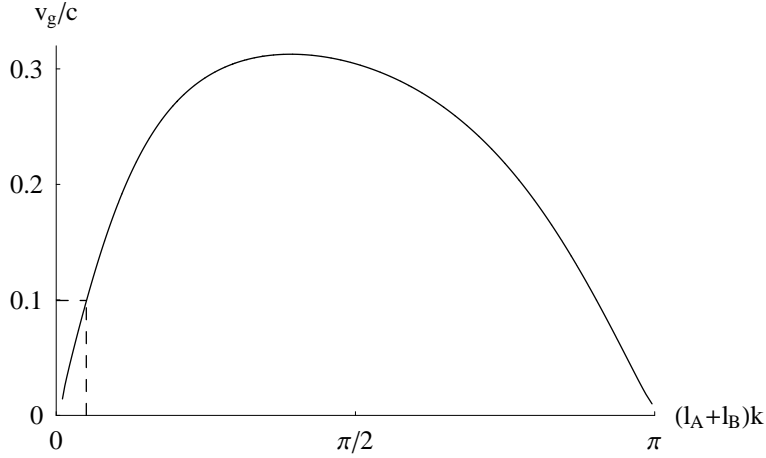}
\end{center}
\caption{The group velocity $v_{g}(k)/c$ of the fourth conduction band
in Fig.~\ref{Ph-BandGap}.
When we have $(l_{A}+l_{B})k=0.158$, we obtain $v_{g}/c=0.0995$.
Broken lines indicates this point.}
\lab{PhC-GroupVel}
\end{figure}

Here, we substitute physical quantities introduced
in Sec.~\ref{section-outline-NS-gate}
into Eq.~(\ref{equation-light-dispersion-relation}).
The implementation is as follows:
The materials $A$ and $B$ are air and GaAs/GaAlAs MQW structure, respectively.
The dielectric constants of them are given by
$\epsilon_{A}/\epsilon_{0}=1.0$ and $\epsilon_{B}/\epsilon_{0}=13.0$.
We assume the thicknesses of layers that compose the photonic crystal are given
by $l_{A}=l_{B}=3.57\times 10^{-7}$ [m].
Using Eqs.~(\ref{wave-vector-I-II})
and (\ref{equation-light-dispersion-relation}),
we obtain the light dispersion relation drawn in Fig.~\ref{Ph-BandGap}.
The group velocity of the light in the photonic crystal can be written in the form
\beq
v_{g}(k_{0})
=
\left.
\frac{d\omega}{dk}
\right|_{k_{0}}.
\lab{definition-group-velocity}
\eeq
We draw the group velocity $v_{g}(k)$ of the fourth conduction band
in Fig.~\ref{PhC-GroupVel}.

Because the variation of $v_{g}$ against $k$ is very steep in the region
where $v_{g}$ takes a small value as shown in Fig.~\ref{PhC-GroupVel},
we need to tune up the wavelength of injected photons precisely
for obtaining the slow group velocity.
(At the same time, we have to adjust $l_{A}$ and $l_{B}$ precisely
because $v_{g}$ is sensitive to the thickness of the layers, as well.)

If we inject photons with $\lambda=8.5\times 10^{-7}$ [m]
and $\omega=2.22\times 10^{15}$ [rad/s] into the photonic crystal,
its wave vector and group velocity are given by
$(l_{A}+l_{B})k=0.295$ and $v_{g}/c=0.171$, respectively.
On the other hand,
if we inject photons with $\lambda=8.47\times 10^{-7}$ [m]
and $\omega=2.23\times 10^{15}$ [rad/s] into the photonic crystal
[this corresponds with $\omega(l_{A}+l_{B})/2\pi c=0.843$],
its wave vector and group velocity are given by
$(l_{A}+l_{B})k=0.158$ and $v_{g}/c=0.0995$, respectively.

Now, we obtain the group velocity $v_{g}$ of photons in the photonic crystal.
However, photons injected into the photonic crystal undergo Kerr nonlinear interaction
only while they travel through the layers of the material $B$.
Thus, we have to estimate the probability $P_{A}$ that the photons stay
in the layers of the material $A$ and
the probability $P_{B}(=1-P_{A})$ that the photons stay in the layers of the material $B$.
Because the number of photons in a certain box is in proportion to energy contained in it,
we obtain the relation
\beqa
P_{A}:P_{B}
&=&
\frac{\omega}{2\pi}
\int_{0}^{2\pi/\omega}dt\:
S\int_{0}^{l_{A}}dz\:
\frac{1}{2}
[\epsilon_{A}E_{\mbox{\scriptsize I}}(z,t)^{2}
+\frac{1}{\mu_{0}}B_{\mbox{\scriptsize I}}(z,t)^{2}]: \non \\
&&\quad\quad
\frac{\omega}{2\pi}
\int_{0}^{2\pi/\omega}dt\:
S\int_{l_{A}}^{l_{A}+l_{B}}dz\:
\frac{1}{2}
[\epsilon_{B}E_{\mbox{\scriptsize II}}(z,t)^{2}
+\frac{1}{\mu_{0}}B_{\mbox{\scriptsize II}}(z,t)^{2}] \non \\
&=&
\int_{0}^{2\pi/\omega}dt\:
\int_{0}^{l_{A}}dz\:
[\frac{\epsilon_{A}}{\epsilon_{0}c^{2}}E_{\mbox{\scriptsize I}}(z,t)^{2}
+B_{\mbox{\scriptsize I}}(z,t)^{2}]: \non \\
&&\quad\quad
\int_{0}^{2\pi/\omega}dt\:
\int_{l_{A}}^{l_{A}+l_{B}}dz\:
[\frac{\epsilon_{B}}{\epsilon_{0}c^{2}}E_{\mbox{\scriptsize II}}(z,t)^{2}
+B_{\mbox{\scriptsize II}}(z,t)^{2}],
\lab{proportion-PA-PB}
\eeqa
where $E_{j}(z,t)$ and $B_{j}(z,t)$ are given by
\beqa
E_{j}(z,t)
&=&
\mbox{Re}
\{[C_{j+}\exp(iK_{j}z)+C_{j-}\exp(-iK_{j}z)]\exp(-i\omega t)\}, \non \\
B_{j}(z,t)
&=&
\mbox{Re}
\{\frac{K_{j}}{\omega}
[C_{j+}\exp(iK_{j}z)-C_{j-}\exp(-iK_{j}z)]\exp(-i\omega t)\} \non \\
&&\quad\quad\mbox{for $j=\mbox{I},\mbox{II}$}.
\eeqa
[We take an average of the energy over time $t$ in Eq.~(\ref{proportion-PA-PB}).]

Substituting $\omega(l_{A}+l_{B})/2\pi c=0.843$ that corresponds
with $\lambda=8.47\times 10^{-7}$ [m]
and $(l_{A}+l_{B})k=0.158$ into the matrix $M$ defined in Eq.~(\ref{definition-matrix-M}),
we can obtain the coefficients of the electromagnetic field,
$C_{\mbox{\scriptsize I}+}$, $C_{\mbox{\scriptsize I}-}$,
$C_{\mbox{\scriptsize II}+}$, and $C_{\mbox{\scriptsize II}-}$
form Eq.~(\ref{Matrix-equation-continuity-condition}).
After carrying out slightly tough numerical calculations,
we obtain
\beq
P_{A}:P_{B}
\simeq
1:11.2.
\eeq
In Sec.~\ref{section-outline-NS-gate}, we obtain the time-of-flight
$\tau_{\mbox{\scriptsize tof}}=\pi/(2\chi)=1.14\times 10^{-10}$ [s]
that we need to flip the sign of the phase in the homogeneous GaAs/GaAlAs MQW material.
Thus, the length of the photonic crystal for realizing the NS gate is given by
\beq
L_{\mbox{\scriptsize PhC}}
=
\tau_{\mbox{\scriptsize tof}}\frac{P_{A}+P_{B}}{P_{B}}v_{g}
=
3.71\times 10^{-3} \mbox{ [m]}.
\lab{whole-length-NS-gate}
\eeq

The photonic crystal with the length
$L_{\mbox{\scriptsize PhC}}=3.71\times 10^{-3}$ [m]
consists of about $10\mbox{ }400$ layers (about $5200$ periods)
of materials $A$ and $B$ (the air and the GaAs/GaAlAs MQW structure).
Moreover, each layer of the GaAs/GaAlAs MQW structure consists of about $15$ periods
of GaAs and GaAs/GaAlAs
(the width of each quantum well is equal to about 100-{\AA}).

In the above estimation, we cannot obtain a great reduction in the length of the device
by the slow group velocity in the photonic crystal.
(The lengths of the homogeneous sample of the Kerr medium and the sample of the Kerr-nonlinear
photonic crystal for the phase rotation are equal to $9.49$ [mm] and $3.71$ [mm],
respectively.)
To obtain a significant reduction, we have to adjust the wavelength of the injected photons
more precisely.
However, as shown in Fig.~\ref{PhC-GroupVel},
the variation of the group velocity $v_{g}$ against the wave vector $k$ is
very steep when $v_{g}$ takes a small value,
so that an accurate adjustment of the wavelength of the injected photons is very difficult.
This is a weak point of our proposition.
However, the author thinks that we can expect the great reduction in the length of the device
by a precise tuning up of the parameters. 

\section{Discussion}
\lab{section-discussion}
According to our estimation, to realize the NS gate,
we have to construct the photonic crystal from about ten thousands layers,
each of whose thickness is equal to about $0.4$ $\mu$m.
Moreover, the layers of the Kerr medium has the GaAs/GaAlAs MQW structure.
These requirements seem to be very severe.
However, for example,
Noda and his collaborators fabricate woodpile-structure of GaAs,
whose typical period is equal to 0.7 $\mu$m
\cite{Noda-1,Noda-2}.
Hence, the author believes that we can construct the NS gate from the one-dimensional
Kerr-nonlinear photonic crystal in the not-too-distant future.

The variation of the group velocity $v_{g}$ against the wave vector $k$ is very steep
in the region where $v_{g}$ takes a small value as shown in Fig.~\ref{PhC-GroupVel}.
Thus, $v_{g}$ is sensitive to the thickness of the layers that compose the photonic crystal
and the wavelength of injected photons,
so that we have to adjust them precisely.
In this paper, we show a plan for the NS gate with the GaAs/GaAlAs MQW structure
as a concrete example.
In our example, we cannot obtain a significant reduction in the length of the device
caused by the slow group velocity in the photonic crystal,
compared to the homogeneous sample of the Kerr medium.
However, the author believes that careful tuning up of the parameters in the laboratory
realizes a feasible design of the device.
Because the adjustment of the physical quantities is very subtle,
we may find another design that is better than the plan we have shown in this paper.

\bigskip
\noindent
{\bf \large Acknowledgment}
\smallskip

\noindent
The author thanks Osamu Hirota for encouragement.

\appendix
\section{The third-order nonlinear susceptibility and the nonlinear refraction coefficient}
\lab{section-third-order-nonlinear-susceptibility-GaAs/GaAlAs}
In this section,
we explain the relation between $\chi^{(3)}$
(the third-order nonlinear susceptibility)
and $n_{2}$ (the nonlinear refraction coefficient),
which is obtained by an experiment for a Kerr medium direct.
(Details of this topic can be found in Ref.~\cite{Yariv}.)
Using this relation, we can calculate $\chi^{(3)}$ from an experimental data of $n_{2}$.

In general,
a nonlinear dielectric polarization of a material is given by
a function of the electric field as
\beqa
P_{i}
&=&
\sum_{j}\chi^{(1)}_{ij}E_{j}
+
\sum_{jk}\chi^{(2)}_{ijk}E_{j}E_{k}
+
\sum_{jkl}\chi^{(3)}_{ijkl}E_{j}E_{k}E_{l}
+... \non \\
&&\quad\quad
\mbox{for $i,j,k,l\in\{x,y,z\}$},
\lab{definition-nonlinear-dielectric-polarization}
\eeqa
where $\chi^{(n)}$
represents the tensor of the $n$th-order nonlinear susceptibility.
Let us assume that an electromagnetic plane wave propagates in the positive $z$-direction
in the material,
so that the electric field $\mbox{\boldmath $E$}$
and the magnetic flux density $\mbox{\boldmath $B$}$
lie in the $x$-$y$ plane.
Letting the $x$ axis and the $y$-axis be parallel to
$\mbox{\boldmath $E$}$ and $\mbox{\boldmath $B$}$
respectively,
we can write the electric field as $\mbox{\boldmath $E$}=(E,0,0)$.
Moreover, to simplify $\mbox{\boldmath $P$}$ defined
in Eq.~(\ref{definition-nonlinear-dielectric-polarization}),
we make the following assumptions about $\chi^{(n)}$:
\beqa
\chi^{(1)}_{ij}
&=&
\delta_{ij}\chi^{(1)} \non \\
\chi^{(2)}_{ijk}
&=&
0
\quad\mbox{for $\forall i,j,k$} \non \\
\chi^{(3)}_{ijkl}
&=&
\left\{
\begin{array}{ll}
\chi^{(3)}_{i} & i=j=k=l \\
0              & \mbox{others}
\end{array}
\right.,
\lab{definition-nonlinear-susceptibilities}
\eeqa
and $\chi^{(n)}_{ij...}=0$ for $n\geq 4$.
Moreover, we write $\chi_{x}^{(3)}=\chi^{(3)}$.

From the above assumptions,
we obtain the dielectric polarization $\mbox{\boldmath $P$}=(P,0,0)$,
where
\beq
P=\chi^{(1)}E+\chi^{(3)}E^{3}
\eeq
Thus, the electric flux density is given by $\mbox{\boldmath $D$}=(D,0,0)$
and
\beq
D
=\epsilon_{0}E+P
=(\epsilon_{0}+\chi^{(1)})E+\chi^{(3)}E^{3},
\eeq
where $\epsilon_{0}$ is the dielectric constant of vacuum.
We define the nonlinear dielectric constant of the material as
\beq
\epsilon(E)
=
\frac{D}{E}
=\epsilon_{0}+\chi^{(1)}+\chi^{(3)}E^{2}.
\eeq
[In general,
we regard $\epsilon(0)[=\epsilon_{0}+\chi^{(1)}]$ as the dielectric constant of the material.]

The refractive index of the matter is given by
\beq
n=\sqrt{\frac{\epsilon\mu}{\epsilon_{0}\mu_{0}}},
\lab{definition-refractive-index}
\eeq
where $\mu_{0}$ and $\mu$ are the magnetic permeabilities of vacuum and the material,
respectively.

Assuming $|\epsilon_{0}+\chi^{(1)}|\gg|\chi^{(3)}E^{2}|$ and $\mu\simeq\mu_{0}$,
which is satisfied for almost all the materials,
we can expand $n$ of Eq.~(\ref{definition-refractive-index})
in powers of $\chi^{(3)}$ as
\beq
n=n_{0}+n'_{2}E^{2}+...,
\lab{refractive-index-expansion-with-E}
\eeq
where
\beqa
n_{0}&=&\sqrt{1+\frac{\chi^{(1)}}{\epsilon_{0}}}, \non \\
n'_{2}&=&\frac{\chi^{(3)}}{2n_{0}\epsilon_{0}}.
\eeqa
However,
because the nonlinear refraction index $n$ is obtained as a function of the field intensity
$I=(1/2)\sqrt{\epsilon/\mu}E^{2}$ in the experiment,
we rewrite Eq.~(\ref{refractive-index-expansion-with-E}) in the form of
\beq
n=n_{0}+n_{2}I,
\eeq
where
\beq
n_{2}
=
\frac{\chi^{(3)}}{n_{0}\epsilon_{0}}\sqrt{\frac{\mu}{\epsilon}}
\simeq
\frac{\chi^{(3)}}{n_{0}^{2}\epsilon_{0}^{2}c},
\lab{nonlinear-refraction-coefficient}
\eeq
and $c=1/\sqrt{\mu_{0}\epsilon_{0}}$ is the light velocity in vacuum.

In Refs.~\cite{Miller-Chemla-Eilenberger-Smith-Gossard-Tsang,
Miller-Chemla-Eilenberger-Smith-Gossard-Wiegmann,
Chemla-Miller-Smith},
nonlinear optical properties of GaAs/GaAlAs multiple quantum well (MQW) material are studied.
In Ref.~\cite{Miller-Chemla-Eilenberger-Smith-Gossard-Wiegmann},
Miller {\it et al}. prepare a sample that consists of
84 periods of 144-{\AA} GaAs and 102-{\AA} GaAs/$\mbox{Ga}_{0.7}\mbox{Al}_{0.3}$As
and obtain the nonlinear refraction coefficient
$n_{2}\sim 1.2\times 10^{-4}$ [$\mbox{cm}^{2}$/W]
at wavelength ($\sim 850$ nm).
Substituting this result of the experiment into Eq.~(\ref{nonlinear-refraction-coefficient}),
we obtain
$\chi^{(3)}=2.97\times 10^{-2}$ (esu)$=3.67\times 10^{-21}$ [m$\cdot$C/$\mbox{V}^{3}$] (SI unit).
(The value of $\chi^{(3)}$ is often represented in cgs-esu unit.
We have the convenient relation,
$\chi^{(3)}_{\mbox{\scriptsize esu}}
=8.1\times 10^{18}\times\chi^{(3)}_{\mbox{\scriptsize SI}}$.)
In the above calculations,
we use the following facts:
The dielectric constant of vacuum is given by $\epsilon_{0}=8.85\times 10^{-12}$ [F/m].
Both the dielectric constants of GaAs and GaAlAs are
given by $\epsilon=13.0\times \epsilon_{0}$.
The magnetic permeabilities of GaAs and GaAlAs are given by $\mu\simeq\mu_{0}$,
where $\mu_{0}$ is the magnetic permeability of vacuum.
Thus, we obtain $n_{0}\simeq\sqrt{13.0}$.
The light velocity in vacuum is given by $c=3.00\times 10^{8}$ [m/s].

The value of $|\chi^{(3)}|$ for GaAs/GaAlAs MQW structure given above is
much larger than those of other materials.
[For example, $\chi^{(3)}$ of C$\mbox{S}_{2}$,
which is a typical Kerr medium, is given by $3.9\times 10^{-13}$ (esu).]

\section{The wave equation and the effective Hamiltonian of the electromagnetic field
in the nonlinear dielectric medium}
\lab{section-wave-equation-effective-Hamiltonian-nonlinear-dielectric-medium}
In this section,
we derive the wave equation and the effective Hamiltonian
of the electromagnetic field in the nonlinear dielectric medium.

First, we consider the wave equation of the electromagnetic field
in the nonlinear dielectric medium.
We use this equation for deriving the photonic band gap
in Sec.~\ref{section-Dispersion-relation-Kerr-nonlinear-photonic-crystal}.
We start from Maxwell's equations in a matter,
where there are no currents and no charges,
\beqa
&&\nabla\cdot\mbox{\boldmath $D$}(\mbox{\boldmath $x$},t)=0, \lab{Maxwell-eq-1} \\
&&\nabla\cdot\mbox{\boldmath $B$}(\mbox{\boldmath $x$},t)=0, \lab{Maxwell-eq-2} \\
&&\nabla\times\mbox{\boldmath $H$}(\mbox{\boldmath $x$},t)
-\frac{\partial}{\partial t}\mbox{\boldmath $D$}(\mbox{\boldmath $x$},t)=0, \lab{Maxwell-eq-3} \\
&&\nabla\times\mbox{\boldmath $E$}(\mbox{\boldmath $x$},t)
+\frac{\partial}{\partial t}\mbox{\boldmath $B$}(\mbox{\boldmath $x$},t)=0. \lab{Maxwell-eq-4}
\eeqa
In these four equations,
$\mbox{\boldmath $D$}$ and $\mbox{\boldmath $B$}$ represent
the electric and the magnetic flux densities,
and
$\mbox{\boldmath $E$}$ and $\mbox{\boldmath $H$}$ represent
the electric and the magnetic fields, respectively.
Moreover, we assume the following relations:
\beqa
\mbox{\boldmath $D$}(\mbox{\boldmath $x$},t)
&=&
\epsilon_{0}\mbox{\boldmath $E$}(\mbox{\boldmath $x$},t)
+
\mbox{\boldmath $P$}(\mbox{\boldmath $x$},t), \non \\
\mbox{\boldmath $B$}(\mbox{\boldmath $x$},t)
&=&
\mu(\mbox{\boldmath $x$})\mbox{\boldmath $H$}(\mbox{\boldmath $x$},t),
\eeqa
where $\mbox{\boldmath $P$}$ represents the dielectric polarization,
$\epsilon_{0}$ represents the dielectric constant of vacuum,
and $\mu$ represents the magnetic permeability
of the material.
Here, we assume $\mu(\mbox{\boldmath $x$})=\mu_{0}$,
where $\mu_{0}$ is the magnetic permeability of vacuum,
because almost all the materials satisfy it.
We also assume that the matter is uniform in the $x$-$y$ plane.

Let us consider the electromagnetic wave that propagates in the positive $z$-direction.
We can describe $\mbox{\boldmath $E$}$ and $\mbox{\boldmath $B$}$
as functions of $z$ and $t$.
Because the matter is uniform in the $x$-$y$ plane,
we can describe $\mbox{\boldmath $P$}$ as a function of $z$ and $t$, as well.

Using Eqs.~(\ref{Maxwell-eq-1}), (\ref{Maxwell-eq-2}),
(\ref{Maxwell-eq-3}), (\ref{Maxwell-eq-4}) and the above assumptions,
we can obtain the following relations:
\beqa
&&\frac{\partial}{\partial z}(\epsilon_{0}E_{z}+P_{z})=0, \lab{Maxwell-eq-a} \\
&&\frac{\partial}{\partial z}B_{z}=0, \lab{Maxwell-eq-b} \\
&&\frac{\partial}{\partial z}B_{y}
+\mu_{0}\frac{\partial}{\partial t}(\epsilon_{0}E_{x}+P_{x})=0, \lab{Maxwell-eq-c} \\
&&\frac{\partial}{\partial z}B_{x}
-\mu_{0}\frac{\partial}{\partial t}(\epsilon_{0}E_{y}+P_{y})=0, \lab{Maxwell-eq-d} \\
&&\frac{\partial}{\partial t}(\epsilon_{0}E_{z}+P_{z})=0, \lab{Maxwell-eq-e} \\
&&\frac{\partial}{\partial z}E_{y}-\frac{\partial}{\partial t}B_{x}=0, \lab{Maxwell-eq-f} \\
&&\frac{\partial}{\partial z}E_{x}+\frac{\partial}{\partial t}B_{y}=0, \lab{Maxwell-eq-g} \\
&&\frac{\partial}{\partial t}B_{z}=0. \lab{Maxwell-eq-h}
\eeqa
From Eqs.~(\ref{Maxwell-eq-a}) and (\ref{Maxwell-eq-e}),
we can take
\beq
\epsilon_{0}E_{z}(z,t)+P_{z}(z,t)=0.
\eeq
In a similar way, from Eqs.~(\ref{Maxwell-eq-b}) and (\ref{Maxwell-eq-h}), we can take
\beq
B_{z}(z,t)=0.
\eeq
Here, we assume that the material is a Kerr medium whose nonlinear susceptibilities
are given by Eq.~(\ref{definition-nonlinear-susceptibilities}).
Thus, the dielectric polarization is given by
\beq
P_{i}=\chi^{(1)}(z)E_{i}+\chi^{(3)}_{i}(z)E_{i}^{3},
\eeq
(we assume that $\chi^{(1)}$ and $\chi^{(3)}_{i}$ do not depend on $t$),
and we obtain
\beq
E_{z}(z,t)=0.
\eeq

Now, $\mbox{\boldmath $E$}(z,t)$ and $\mbox{\boldmath $B$}(z,t)$ lie in the $x$-$y$ plane.
Letting the $x$-axis be parallel with $\mbox{\boldmath $E$}(z,t)$,
we can write $\mbox{\boldmath $E$}(z,t)=(E(z,t),0,0)$.
Thus, from Eqs.~(\ref{Maxwell-eq-d}) and (\ref{Maxwell-eq-f}),
we obtain $\mbox{\boldmath $B$}(z,t)=(0,B(z,t),0)$.
From Eqs.~(\ref{Maxwell-eq-c}) and (\ref{Maxwell-eq-g}),
we obtain the wave equations for $E(z,t)$ and $B(z,t)$,
\beqa
&&
\frac{\partial^{2}}{\partial z^{2}}E(z,t)
-
\mu_{0}\epsilon(z)\frac{\partial^{2}}{\partial t^{2}}E(z,t) \non \\
&&\quad\quad
-
\mu_{0}\chi^{(3)}(z)\frac{\partial^{2}}{\partial t^{2}}E(z,t)^{3}=0,
\lab{wave-equation-electric-field}
\eeqa
\beq
\frac{\partial}{\partial z}E(z,t)
=
-\frac{\partial}{\partial t}B(z,t),
\lab{wave-equation-magnetic-field}
\eeq
where $\epsilon(z)=\epsilon_{0}+\chi^{(1)}(z)$
and $\chi^{(3)}=\chi^{(3)}_{x}$.

Second, we consider the effective Hamiltonian of the electromagnetic field
in a homogeneous Kerr medium.
We assume that the electromagnetic field propagates in the positive $z$-direction,
and it is described by
Eqs.~(\ref{wave-equation-electric-field}) and (\ref{wave-equation-magnetic-field}).
Because the material is uniform in all directions,
we can regard $\chi^{(1)}$, $\chi^{(3)}$, and $\epsilon$ as constants.

In general,
the energy of the electromagnetic field confined in the box of volume $\mbox{Vol}$ is given by
\beq
\mathcal{H}
=
\int_{\mbox{\scriptsize Vol}}d^{3}\mbox{\boldmath $x$}\:
\frac{1}{2}
(\mbox{\boldmath $D$}\cdot\mbox{\boldmath $E$}+\mbox{\boldmath $H$}\cdot\mbox{\boldmath $B$}).
\lab{EM-field-energy-density}
\eeq
Because we make the following assumptions now,
$\mbox{\boldmath $E$}=(E(z,t),0,0)$,
$\mbox{\boldmath $D$}=(\epsilon E(z,t)+\chi^{(3)}E(z,t)^{3},0,0)$,
$\mbox{\boldmath $B$}=(0,B(z,t),0)$,
$\mbox{\boldmath $H$}=(0,(1/\mu_{0})B(z,t),0)$,
we can rewrite $\mathcal{H}$ given in Eq.~(\ref{EM-field-energy-density}) as
\beqa
\mathcal{H}
&=&
\mathcal{H}_{0}+V, \non \\
\mathcal{H}_{0}
&=&
\int_{\mbox{\scriptsize Vol}}d^{3}\mbox{\boldmath $x$}\:
\frac{1}{2}
(\epsilon E^{2}
+
\frac{1}{\mu_{0}}B^{2}), \non \\
V
&=&
\int_{\mbox{\scriptsize Vol}}d^{3}\mbox{\boldmath $x$}\:
\frac{\chi^{(3)}}{2}E^{4}.
\eeqa

From now on,
we perform the quantization of the free electromagnetic field
whose Hamiltonian is given by $\mathcal{H}_{0}$.
We regard $V$ as a perturbation,
and we construct the effective Hamiltonian of the quantized field
by evaluating the first-order perturbed energy.
For the quantization of the free field,
we write $E(z,t)$ and $B(z,t)$ as products:
\beqa
E(z,t)
&=&
-\frac{1}{\sqrt{\epsilon}}\sum_{l}
\dot{q}_{l}(t)u_{l}(z), \lab{product-form-E} \\
B(z,t)
&=&
\frac{1}{\sqrt{\epsilon}}\sum_{l}
q_{l}(t)u_{l}'(z). \lab{product-form-B}
\eeqa
The free electric field $E(z,t)$ has to satisfy Eq.~(\ref{wave-equation-electric-field})
with $\chi^{(3)}=0$.
Assuming the normalization of $u_{l}(z)$ in volume $\mbox{Vol}=Sd$
and the boundary condition $u_{l}(0)=u_{l}(d)=0$,
we obtain
\beq
u_{l}(z)=\sqrt{\frac{2}{Sd}}\sin\frac{l\pi z}{d}
\quad\quad
\mbox{for $l=\pm 1,\pm 2,...$}.
\lab{special-solutions-u-l}
\eeq
Substituting Eqs.~(\ref{product-form-E}), (\ref{product-form-B}),
and (\ref{special-solutions-u-l}) into $\mathcal{H}_{0}$,
we obtain
\beqa
\mathcal{H}_{0}
&=&
\sum_{l=\pm 1,\pm 2,...}\mathcal{H}_{l}, \\
\mathcal{H}_{l}
&=&
\frac{1}{2}(\dot{q}_{l}^{2}+\omega_{l}^{2}q_{l}^{2}),
\eeqa
where
\beq
\omega_{l}=\sqrt{\frac{1}{\mu_{0}\epsilon}}\frac{|l|\pi}{d}.
\eeq

Let us define a new variable $p_{l}\equiv\dot{q}_{l}$.
Then, we can rewrite $\mathcal{H}_{l}$ as
$\mathcal{H}_{l}=(1/2)(p_{l}^{2}+\omega_{l}^{2}q_{l}^{2})$,
and we obtain the relations
$\partial \mathcal{H}_{l}/\partial q_{l}=-\dot{p}_{l}$
and
$\partial \mathcal{H}_{l}/\partial p_{l}=\dot{q}_{l}$.
Thus, we can regard $q_{l}$ and $p_{l}$ as canonical conjugate variables.
We perform the quantization of the field by taking the commutation relation
$[q_{l},p_{m}]=i\hbar\delta_{lm}$.
This commutation relation corresponds with the following relations:
\beqa
q_{l}&=&\sqrt{\frac{\hbar}{2\omega_{l}}}(a_{l}^{\dagger}+a_{l}), \\
p_{l}&=&i\sqrt{\frac{\hbar\omega_{l}}{2}}(a_{l}^{\dagger}-a_{l}), \lab{canonical-variable-p-l}
\eeqa
$[a_{l},a_{m}^{\dagger}]=\delta_{lm}$, and
$[a_{l},a_{m}]=[a_{l}^{\dagger},a_{m}^{\dagger}]=0$.
Rewriting $\mathcal{H}_{l}$ with the creation and annihilation operators
$a_{l}^{\dagger}$ and $a_{l}$ as
\beq
\mathcal{H}_{l}=\hbar\omega_{l}(a_{l}^{\dagger}a_{l}+\frac{1}{2}),
\eeq
we accomplish the quantization of the free field.

Next, we derive the effective Hamiltonian of the quantized field
for $V$ as the first-order perturbed energy.
We substitute Eqs.~(\ref{product-form-E}), (\ref{special-solutions-u-l}),
and (\ref{canonical-variable-p-l}) into $V$,
and we obtain
\beqa
V
&=&
\frac{1}{2}\chi^{(3)}\frac{1}{\epsilon^{2}}(\frac{2}{Sd})^{2}
\sum_{k,l,m,n=\pm 1,\pm 2,...}
p_{k}p_{l}p_{m}p_{n} \non \\
&&\quad\quad\times
\int_{Sd}d^{3}\mbox{\boldmath $x$}\:
\sin\frac{k\pi z}{d}
\sin\frac{l\pi z}{d}
\sin\frac{m\pi z}{d}
\sin\frac{n\pi z}{d}.
\eeqa
Using the following formula,
\beqa
&&\int_{0}^{d}dz
\sin\frac{k\pi z}{d}
\sin\frac{l\pi z}{d}
\sin\frac{m\pi z}{d}
\sin\frac{n\pi z}{d} \non \\
&=&
\frac{d}{8}
(
-\delta_{k,l+m+n}
+\delta_{k,-l+m+n}
+\delta_{k,l-m+n}
-\delta_{k,-l-m+n} \non \\
&&\quad\quad
+\delta_{k,l+m-n}
-\delta_{k,-l+m-n}
-\delta_{k,l-m-n}
+\delta_{k,-l-m-n}
),
\eeqa
we obtain
\beqa
V
&=&
\frac{\chi^{(3)}}{4\epsilon^{2}Sd}
\sum_{k,l,m,n=\pm 1,\pm 2,...}
p_{k}p_{l}p_{m}p_{n} \non \\
&&\quad\quad\times
(-4\delta_{k,l+m+n}+3\delta_{k+l,m+n}+\delta_{k+l+m+n,0}).
\eeqa

Here, we assume that we inject only photons of a certain mode $l$ into the material.
Thus, terms that contain $a_{k}$ ($k\neq l$) vanish when we apply them to the state vector.
Moreover, we neglect events where photons of mode $k(\neq l)$ are created.
Thus, for example, we ignore the case where three photons of $\omega_{l}$ are
annihilated and one photon of $\omega_{3l}(=3\omega_{l})$ is created.
This implies that the effective terms of $V$ cannot contain
$a_{k}^{\dagger}$ ($k\neq l$), as well.
Hence, the effective potential for photons of mode $l$ is given by
\beq
\frac{3}{4}
\frac{\chi^{(3)}}{\epsilon^{2}Sd}
(\frac{\hbar\omega_{l}}{2})^{2}(a_{l}^{\dagger}-a_{l})^{4}.
\eeq

From some calculations, we obtain
\beqa
&&
(a_{l}^{\dagger}-a_{l})^{4} \non \\
&=&
6(a_{l}^{\dagger})^{2}a_{l}^{2}
+8a_{l}^{\dagger}a_{l}
+1 \non \\
&&\quad
+(a_{l}^{\dagger})^{4}+a_{l}^{4}
-4[(a_{l}^{\dagger})^{2}+a_{l}^{2}
+(a_{l}^{\dagger})^{3}a_{l}
+a_{l}^{\dagger}a_{l}^{3}].
\eeqa
Here, we concentrate on the terms which conserve the number of photons.
Hence, we can write the effective Hamiltonian as follows:
\beqa
\mathcal{H}
&=&
\hbar\omega_{l}(a_{l}^{\dagger}a_{l}+\frac{1}{2}) \non \\
&&\quad
+
\frac{3}{4}
\frac{\chi^{(3)}}{\epsilon^{2}Sd}
(\frac{\hbar\omega_{l}}{2})^{2}
[6(a_{l}^{\dagger})^{2}a_{l}^{2}
+8a_{l}^{\dagger}a_{l}
+1] \non \\
&=&
\tilde{\mathcal{H}}_{l}+\tilde{V}_{l},
\eeqa
and
\beqa
\tilde{\mathcal{H}}_{l}
&=&
\hbar\omega_{l}(a_{l}^{\dagger}a_{l}+\frac{1}{2}) \non \\
&&\quad
+
\frac{3}{4}
\frac{\chi^{(3)}}{\epsilon^{2}Sd}
(\frac{\hbar\omega_{l}}{2})^{2}
[8a_{l}^{\dagger}a_{l}+1], \non \\
\tilde{V}_{l}
&=&
\frac{9}{8}
\frac{\chi^{(3)}\hbar^{2}\omega_{l}^{2}}{\epsilon^{2}Sd}
(a_{l}^{\dagger})^{2}a_{l}^{2}.
\eeqa

When we estimate the phase rotation induced by Kerr nonlinear interaction,
$\tilde{V}_{l}$ makes a main contribution and $\tilde{\mathcal{H}}_{l}$ causes
only a background phase rotation.
Furthermore,
in most cases,
the relation $\omega_{l}\gg|\chi^{(3)}|\hbar\omega_{l}^{2}/(\epsilon^{2}Sd)$
is satisfied and we can take $\tilde{\mathcal{H}}_{l}\simeq\mathcal{H}_{l}$.
Hence, we obtain the following effective Hamiltonian:
\beq
\mathcal{H}=\mathcal{H}_{l}+V_{l},
\eeq
where
\beqa
\mathcal{H}_{l}
&=&
\hbar\omega_{l}(a_{l}^{\dagger}a_{l}+\frac{1}{2}), \non \\
V_{l}
&=&
\frac{9}{8}
\frac{\chi^{(3)}\hbar^{2}\omega_{l}^{2}}{\epsilon^{2}Sd}
(a_{l}^{\dagger})^{2}a_{l}^{2}.
\eeqa

\end{document}